 \documentclass[10pt,journal,compsoc]{IEEEtran}

\ifCLASSOPTIONcompsoc
  \usepackage[nocompress]{cite}
\else
  \usepackage{cite}
\fi

\usepackage{booktabs} 
\usepackage{fancyvrb}
\usepackage{listings}
\usepackage{color}
\usepackage[dvipsnames]{xcolor}
\usepackage{balance}
\usepackage{fancybox}
\usepackage{xspace}
\usepackage{subcaption}
\usepackage{multirow}
\usepackage{url}
\usepackage{array}
\usepackage{graphicx}
\usepackage{bbding}
\usepackage{wasysym}
\usepackage{amssymb}
\usepackage{pdflscape}
\usepackage{tabularx,colortbl}
\usepackage{dirtree}
\usepackage{rotating}

\usepackage{comment}

\usepackage{algorithm}
\usepackage[noend]{algpseudocode}
\usepackage{enumitem}

\usepackage{colortbl}
\usepackage{tabu}
\usepackage[skins]{tcolorbox}
\usepackage{cleveref}

\usepackage{balance}

\usepackage{float}

\floatstyle{ruled}
\newfloat{program}{thp}{lop}
\floatname{program}{Program}

\usepackage{array, booktabs, boldline} %

\ifCLASSINFOpdf
\else
\fi

\definecolor{grayboxcolor}{HTML}{f2f2f2}
\definecolor{highlight_orange}{HTML}{d98b5e}

\newcommand{\conclusion}[1]{%
	\begin{center}\noindent\thicklines\setlength{\fboxsep}{8pt}\fcolorbox{black}{grayboxcolor}{\begin{minipage}{3.3in}\textbf{#1}\end{minipage}}\end{center}} 
 
\newcommand{\highlight}[1]{%
	\vspace{-1mm}\begin{tcolorbox}[enhanced,colback=white,frame hidden,
borderline west={1mm}{0mm}{highlight_orange}]
\vspace{-3mm}\textit{#1}\vspace{-3.5mm}
\end{tcolorbox}}

\newcommand{\rqi}{RQ1: How does dependency management impact the risk of vulnerabilities for downstream packages?}
\newcommand{\rqii}{RQ2: How can we identify packages that quickly mitigate vulnerabilities?}
\newcommand{\rqiii}{RQ3: How do developers perceive dependency practices for vulnerability mitigation?}

\definecolor{cadmiumgreen}{rgb}{0.0, 0.42, 0.24}

\def\commentsenabled{}

\ifdefined\commentsenabled
\newcommand{\diego}[1]{\textcolor{cadmiumgreen}{Diego: #1}} 
\newcommand{\ahmad}[1]{\textcolor{orange}{Ahmad: #1}} 
\newcommand{\abbas}[1]{\textcolor{red}{Abbas: #1}} 
\newcommand{\todo}[1]{\textcolor{red}{TODO: #1}}
\else
\newcommand{\diego}[1]{} 
\newcommand{\ahmad}[1]{} 
\newcommand{\abbas}[1]{} 
\newcommand{\todo}[1]{} %
\fi

\begin{document}
%
\title{Dependency Practices for Vulnerability Mitigation}

%
%
%
%

%

\author{Abbas~Javan~Jafari,
	Diego~Elias~Costa,
	Ahmad~Abdellatif,
        Emad~Shihab,~\IEEEmembership{Senior Member,~IEEE}
\IEEEcompsocitemizethanks{\IEEEcompsocthanksitem Abbas~Javan~Jafari, Ahmad Abdellatif and Emad Shihab are with the Data-driven Analysis of Software (DAS) Lab at the Department of Computer Science and Software Engineering, Concordia University, Montreal, Canada.\protect\\
E-mail: {a\_javanj@encs.concordia.ca,  ahmad.abdellatif@concordia.ca, emad.shihab@concordia.ca}
\protect\\
\IEEEcompsocthanksitem Diego~Elias~Costa is with the Department of Computer Science and Software Engineering, Concordia University, Montreal, Canada. \protect\\
E-mail: diego.costa@concordia.ca
}
\thanks{Manuscript received August 27, 2023; revised December 06, 2024.}}

\IEEEtitleabstractindextext{%
\begin{abstract}
Relying on dependency packages accelerates software development, but it also increases the exposure to security vulnerabilities that may be present in dependencies. While developers have full control over which dependency packages (and which version) they use, they have no control over the dependencies of their dependencies. Such transitive dependencies, which often amount to a greater number than direct dependencies, can become infected with vulnerabilities and put software projects at risk. To mitigate this risk, Practitioners need to select dependencies that respond quickly to vulnerabilities to prevent the propagation of vulnerable code to their project. To identify such dependencies, we analyze more than 450 vulnerabilities in the npm ecosystem to understand why dependent packages remain vulnerable. We identify over 200,000 npm packages that are infected through their dependencies and use 9 features to build a prediction model that identifies packages that quickly adopt the vulnerability fix and prevent further propagation of vulnerabilities. We also study the relationship between these features and the response speed of vulnerable packages. We complement our work with a practitioner survey to understand the applicability of our findings. Developers can incorporate our findings into their dependency management practices to mitigate the impact of vulnerabilities from their dependency supply chain.
\end{abstract}

\begin{IEEEkeywords}
Software ecosystems, Dependency management, Update strategies, npm
\end{IEEEkeywords}}

\maketitle

\IEEEdisplaynontitleabstractindextext

%
\IEEEpeerreviewmaketitle

\section{Introduction}
\label{sec:Introduction}

Software ecosystems have facilitated large scale code reuse by providing access to a wide range of software packages. In turn, developers are increasingly reliant on third-party packages to accelerate development \cite{bombonatti2017synergies}. However, developers also need to expend effort and expertise to manage their dependencies \cite{kula2018developers}. In a recent survey, developers reported that they are not confident in their current dependency management practices \cite{tidelift2022}. The Node Package Manager (npm) is the world's largest software package ecosystem with more than 2 million packages \cite{npmjs}. The average number of packages in the npm ecosystem is growing year after year \cite{sonatype2021}, but more importantly, the number of dependencies per package is growing at a super-linear rate \cite{zimmermann2019small}. The scale and complexity of the npm dependency network has created many dependency management challenges \cite{artho2012software, decan2019empirical, rickard2021, decan2017empirical}. 

Security vulnerabilities are a prevalent issue in software ecosystems. The increase of deep dependency supply chains provides an exploitable opportunity for attackers. In the year 2021, there was a 650\% increase in attacks aimed at exploiting vulnerabilities in upstream software packages \cite{sonatype2021}. Up to 40\% of all npm packages depend on code that is infected with a publicly disclosed vulnerability \cite{zimmermann2019small} and the number of reported vulnerabilities in npm is increasing exponentially \cite{zerouali2022impact}. The interconnected nature of software ecosystems increases the threat surface for vulnerabilities \cite{liu2022demystifying, decan2018impact}. For example, when the popular ``lodash'' package was infected with a high severity vulnerability, more than 4 million open-source projects were exposed to a potential attack \cite{lodash_vul2019}. When a client installs an npm package, they are implicitly trusting up to 80 other packages (on average), many of which are due to transitive dependencies (dependencies of dependencies) \cite{zimmermann2019small}. More than one third of the latest package releases in npm are exposed to vulnerabilities through their transitive dependencies \cite{zerouali2022impact}. Vulnerabilities in transitive dependencies can propagate to our project and yet, we have no direct control over what transitive dependencies (or what version) are installed alongside our project. Long dependency chains created through layers of transitive dependencies also impede the propagation of vulnerability fixes throughout the affected packages in the ecosystem \cite{chinthanet2021lags, alfadel2023discoverability}.


Since transitive dependencies are installed based on the requirements of our direct dependencies, the only means to address our exposure to vulnerabilities in transitive dependencies is relying on the responsiveness of our direct dependencies. \textit{Responsive packages are packages that  install the published fix for their vulnerable dependencies in a timely manner}. Developers need to align their dependency practices to select responsive dependency packages. We aim to identify and understand the attributes of npm packages that indicate better responsiveness to vulnerable dependencies. By selecting packages that quickly adopt vulnerability fixes in their dependencies, developers can reduce the risk of publicly disclosed vulnerabilities from transitive dependencies. Our research is formulated through the following questions:

We extract vulnerability data for 458 vulnerabilities from the npm advisory database. We then cross-reference the vulnerabilities with more than 154 million dependency relationships in the npm ecosystem, extracted from libraries.io to identify 201,027 unique packages that install (and later fix) vulnerable dependencies. We measure how long it takes for the dependent packages to adopt the vulnerability fix. We train a random forest model that uses 9 package attributes (e.g. release frequency) to classify the \textit{fast-reponder} and \textit{slow-responder} packages.


\textit{\rqi} While it is up to the maintainer of a vulnerable package to release a fix, dependent packages must decide on when and how they install the fix in their dependency. The average public vulnerability is fixed in a little over 1, but it takes on average more than 6 months for dependents to install the fixed version, which highlights the importance of \textit{package responsiveness} on mitigating vulnerabilities. Dependency management decisions of packages with a vulnerable dependency also influence what portion of fixes can be installed. By relying only on patch updates, dependents miss out on 30\% (on average) of vulnerability fixes.

\textit{\rqii} Developers need a means to identify and select responsive packages to mitigate the risk of vulnerabilities from transitive dependencies. The high ROC-AUC score of 0.85 for our classification model shows that package attributes are useful indicators for differentiating between fast-responder and slow-responder packages. We found that non-restrictive dependency update strategy have a shorter exposure to vulnerabilities. Package age, and release frequency are also good predictors of how quickly packages adopt a vulnerability fix.

\textit{\rqiii}
In order to gauge the applicability of our results in practice and understand the perception of practitioners on our findings, we design a survey and disseminate it among 67 practitioners from industry and academia. Many practitioners are unaware that the lack of updates from downstream dependents is the key culprit for prolonged exposure to public vulnerabilities. In regards to the package attributes that indicate a faster adoption of vulnerability fixes, the experience of practitioners is generally aligned with our results. Survey participants are in favor of incorporating our findings into their dependency management practices.

This paper is structured as follows. Section~\ref{sec:Background} provides readers with the necessary background for the study. Section~\ref{sec:data-curation} explains the methodology and dataset used to conduct the study. We present the findings of our research in Section~\ref{sec:Results} and discuss the practical and research implications of our results in Section~\ref{sec:Implications}. In Section~\ref{sec:Related Work}, we summarize the key related works. Section~\ref{sec:Threats to Validity} presents the threats to validity. We conclude our study in Section~\ref{sec:Conclusion}.
\section{Background}
\label{sec:Background}
In this section, we explain the background and terminologies needed for our study. We describe the dependency management process in the npm ecosystem and the semantic versioning guidelines.


\begin{figure}[h]
    \centering
    \includegraphics[width=\linewidth]{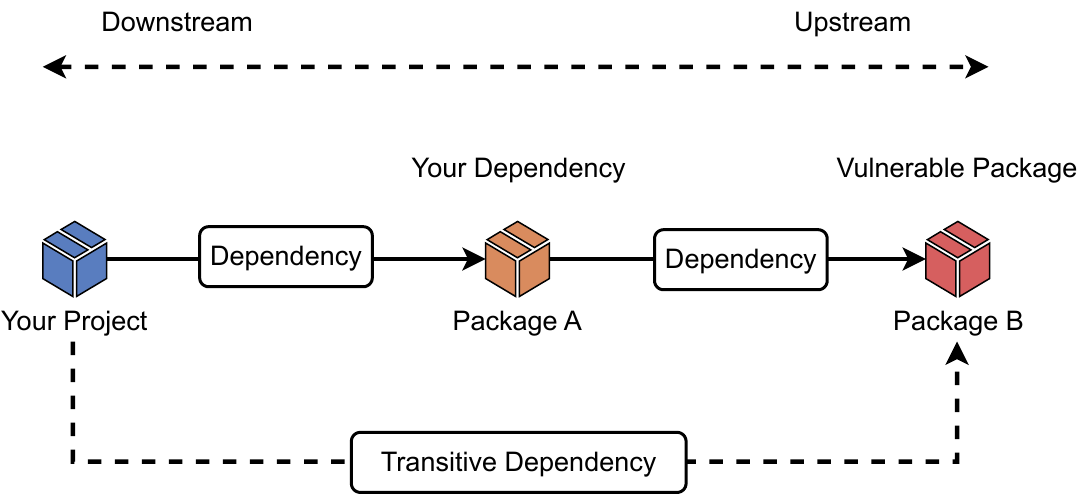}
    \caption{Direct and transitive dependency process}
    \label{fig:dependency_process}
\end{figure}

The npm ecosystem hosts more than 4 million software packages \cite{libraries_npm}. When a package relies on another package for correct functionality, we consider the former as a \textit{dependent package} and the latter as a \textit{dependency package} (Figure~\ref{fig:dependency_process}). However, the dependency package can also rely on other packages for its own functionality. This creates a transitive dependency between the npm package and the dependencies of its dependencies. When considering a package in npm, any dependency package (direct or transitive) is considered an \textit{upstream package} and any dependent package (direct or transitive) is considered a \textit{downstream package}. Since it can be difficult to account for all the upstream packages for any given package, it is uncommon for a single dependency declaration in our project to result in only installing a single package. When installing an npm package, we are implicitly trusting (and affected by) 79 upstream packages on average \cite{zimmermann2019small}. This creates a considerable attack surface for our project should any upstream package becomes infected with a vulnerability.

\begin{figure}[h]
    \centering
    \fbox{\includegraphics[width=0.85\linewidth]{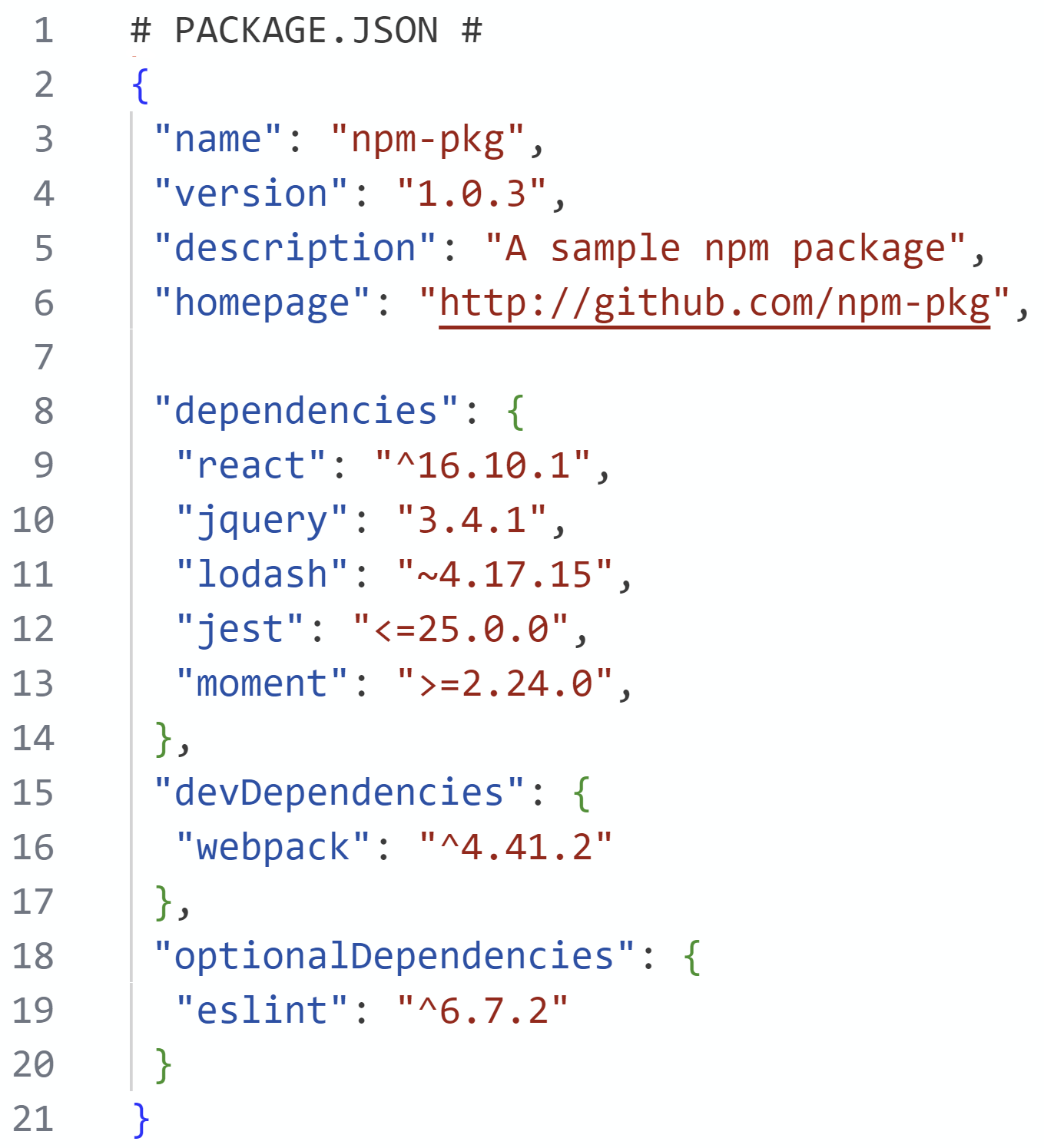}}
    \caption{Dependency configuration file in npm}
    \label{fig:package_json}
\end{figure}

The package.json file is the de facto configuration document for packages in the npm ecosystem. These files include package metadata such as name and version but are also where package maintainers explicitly specify their dependency configuration. Figure~\ref{fig:package_json} depicts an example of a package.json file. In this example, if \textit{jquery} version 3.4.1 becomes infected with a vulnerability which is later fixed in version 3.4.2, the example package will remain vulnerable. The ``dependencies'' object is used to specify runtime dependencies which are required for the package to run. The development dependencies are used during package development and testing and are not required for downstream dependents. In regards to optional dependencies, the package manager will try to fetch them, but no errors will be raised if the package manager is unsuccessful.

When a package is installed using the \textit{npm install} command, the package manager starts installing all the upstream runtime dependencies, which includes both the direct dependencies and transitive dependencies of the package. The process continues until the full dependency tree is installed. Additionally, the package manager generates a package-lock.json which contains the complete list of installed dependencies and their versions. The lock file can be used to trace any changes in the installed versions of packages in the dependency tree and to ensure consistent installations in the future.


The npm ecosystem recommends package maintainers to use Semantic Versioning (SemVer) to manage their releases \cite{npmsemver}. SemVer is a means for package maintainers to communicate the type changes in a new versions so downstream dependents can decide whether they want to update or continue using an older version. SemVer defines a format for version numbering in the form of [major].[minor].[patch]. Any new release that contains backward incompatible feature updates should bump the major version number, a release that contains backward compatible features should bump the minor version number and a release that only contains fixes should bump the patch version number \cite{semver2019}. Downstream dependents can manually install any particular version of a dependency but they can also use pre-defined constraints to allow the package manager to govern how dependencies are updated. Section~\ref{sec:data-curation} explains how these constraints are used by developers.


%
\section{Methodology and Data}
\label{sec:data-curation}
Our objective is to identify and understand the characteristics of responsive packages. Consequently, we require data regarding vulnerable packages in npm. We also need information about the attributes of downstream packages that depend on these vulnerable packages.
We combine the data provided by the npm security advisories and package and dependency metadata for the npm ecosystem to curate a dataset of vulnerable dependency relationships in npm. We then extract the relevant features from the downstream dependents that install (and later fix) a vulnerable version of the upstream dependency to prepare our dataset for training a machine learning model. A complete replication package for this study is available on Zenodo \cite{replication_vul}.

\subsection{Vulnerable dependency dataset}


 In order to analyze the package attributes that indicate a faster adoption of vulnerability fixes, we need first to identify packages that install a vulnerable dependency. We gather the dataset of reviewed npm security advisories from GitHub \cite{github_advisories}.
 The vulnerability advisories include the vulnerability name, published date, severity, CWE classification \cite{cwe}, CVE identifier \cite{cve}, affected versions, fixed versions, along with a description of the vulnerability. We also use the latest available version of the libraries.io dataset \cite{libraries_npm} to identify dependency relationships in the npm ecosystem so we can extract packages that install a vulnerable dependency. This dataset has been used and validated in previous research \cite{decan2019empirical, jafari2023package}. The dataset contains metadata for 1,275,082 unique packages, 11,400,714 package versions and 154,914,774 dependency relationships for the npm ecosystem. We then cross-reference both datasets and select the packages that have a vulnerable version in our npm dependency dataset, resulting in 118, 253, 160 and 44 critical, high, medium and low severity vulnerabilities, respectively. We maintain 4 separate datasets (one for each vulnerability severity type) to facilitate detailed analyses. 


For each vulnerability, we first collect the entire list of its downstream dependents from the dataset to identify the packages that are 'potentially vulnerable' to the upstream vulnerability. We only consider runtime dependency relationships from the downstream dependents to upstream packages because they are required for the package to properly function and should be complete (missing runtime dependencies are considered a bad practice \cite{jafari2020dependency}). This first step amounts to more than 1,940,000, 4,545,000, 4,770,000 and 2,320,000 potentially vulnerable dependencies for critical, high, medium and low severity vulnerabilities, respectively. 
We then evaluate each of the mentioned dependency relationships to determine whether the downstream dependency actually installs a vulnerable version of the upstream package (i.e. downstream dependent installs a vulnerable version).
Similarly, we determine if the downstream package installs a version including the fix or any version higher than that (i.e. downstream dependent installs a non-vulnerable version). Since disclosed vulnerabilities often affect all previous releases of a package \cite{zerouali2022impact}, the most reliable way to mitigate vulnerabilities is to update to a fixed version. 

Many packages specify flexible version ranges for their dependencies which may install a different upstream version depending on the evaluation time frame (i.e. constraint of $>$1.2 may install version 1.5 or version 2.0 depending on when the constraint is evaluated). In order to verify if a package installs a vulnerable version, we evaluate the current downstream constraint the day before the fix could be adopted. If no downstream dependent in our dataset installs the vulnerable upstream version, the upstream package are removed from our analysis. In order to verify if the downstream package installs a fix, we evaluate the current downstream constraint at the time the fix was released by the upstream package and every subsequent date the downstream releases a new version and measure the first time the fixed version (or any higher version) of the upstream package is installed. Table~\ref{tab:dataset_stats} presents the detailed statistics of the final dataset used in our study. Note that total vulnerable dependents are less than the sum of vulnerable dependents for each severity type since a dependent package can appear as a vulnerable dependent in multiple severity type groups.

\begin{table*}[t]
	\centering
	\caption{The dataset for our study}
	\label{tab:dataset_stats}
	



	

\begin{tabular}{lrrrrr}
	\toprule
	\textbf{Dataset Attribute} &  \textbf{Critical Severity} &  \textbf{High Severity}&  \textbf{Medium Severity}&  \textbf{Low Severity} & \textbf{Total}\\
	
	\midrule
	Vulnerabilities 		&	59 &	213&	146&	40&	\textbf{458}\\
	Vulnerable Dependencies 		& 	113,068 & 	165,311& 	151,471& 	79,746& 	\textbf{509,596}\\
	Vulnerable Dependents 		& 	104,149 & 	134,329& 	126,087& 	75,432& 	\textbf{201,027}\\
        \bottomrule
\end{tabular}

\end{table*}

\subsection{Package feature extraction}

We aim to identify and collect features in downstream packages that are relevant for responding to vulnerable dependencies. The features will be used to train a model to predict the speed of response to a vulnerable dependency fix. We aim to consider dimensions regarding package popularity, maturity \& stability, activity and dependency management.



\textbf{Popularity:} 
We use \textit{Dependent Count} as our popularity metric as it is the de facto measure of the number of unique downstream clients for a package. We hypothesize that highly used packages may take greater care in quickly addressing vulnerable dependencies as their vulnerable dependencies can transitively impact a larger number of downstream packages and consequently, invite greater repercussions. Multiple studies highlight the number of downstream dependents as a metric used by developers to aid in their dependency selection \cite{pashchenko2020qualitative,larios2020selecting,bogart2016break,haenni2013categorizing}.

\textbf{Maturity \& Stability:}
We use \textit{Age} and \textit{Release Status} as our indicators for package maturity \& stability. Packages with a longer history provide practitioners with more reliable information regarding the responsiveness of the package to security vulnerabilities. On the other hand, older projects are less likely to still be well maintained (software also rots after all). Developers have cited package maturity as a criterion for selecting dependencies \cite{larios2020selecting}.
The release status determines whether a package has released their first 1.0.0 version or if they are still in their initial development releases (e.g. v0.2.3). SemVer considers pre-1.0.0 releases unstable by nature and we believe their use is a reliable means to gauge the maintainers' perception regarding the maturity of their own package. Release status has previously been referenced by developers as a metric in selecting packages \cite{bogart2016break}.

\textbf{Activity:} 
We use the \textit{Release Frequency} as our metric for measuring activity. We hypothesize that packages with a more active development and maintenance schedule are more proactive in responding to vulnerable dependencies. If a package rarely releases a new version, they need to rely on automatic dependency updates as their means of installing new fixes for vulnerable dependencies. However, a package that limits automatic dependency updates can still install the fix manually if they release a new version of their package with a modified configuration file. Release frequency normalizes the number of total releases by the age of a package and it is an indicator used by developers when selecting dependencies \cite{larios2020selecting,bogart2016break,haenni2013categorizing}. 

\textbf{Dependency Management:} 
Since we want to identify features that best predict the responsiveness to vulnerable dependencies, we need to consider a feature group that focuses on the dependency management of packages. We hypothesize the dependency-related decisions made by a potential direct dependency has a considerable influence on our exposure to transitive dependencies. We use \textit{Dependency Count}, \textit{Dependency Modifications} and \textit{Dependency Update Strategy} as our dependency management features. \textit{Dependency Count} is a measure of the number of dependencies, and it is considered by developers to aid in dependency selection \cite{larios2020selecting, mujahid2023characteristics}. We hypothesize that a larger number of dependencies may make it difficult to properly manage and keep track of vulnerable dependencies. \textit{Dependency Modifications} is a measure of dependency change and measures how many releases modified the dependency configuration. Changing a dependency configuration file (package.json) more frequently can be a sign of continuous upkeep to maintain dependency health or it could be a sign of frequent dependency problems. When viewed alongside release frequency, the number of dependency modifications highlights whether an active package with frequent releases also frequently modifies their dependencies.

\begin{figure}[h]
    \centering
    \includegraphics[width=0.75\linewidth]{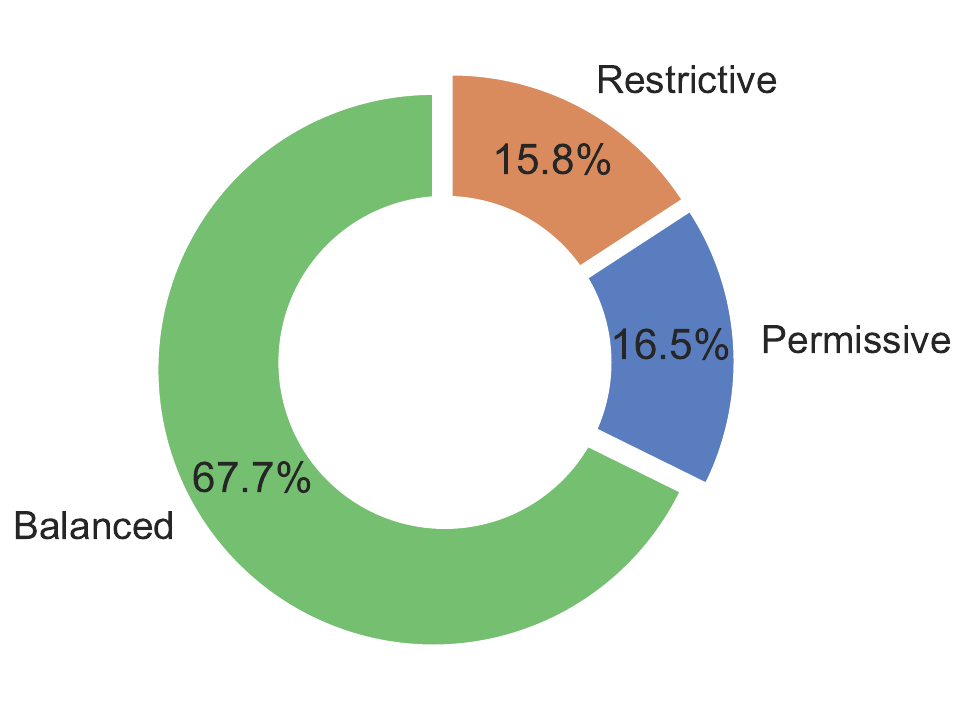}
    \caption{Distribution of the update strategy feature}
    \label{fig:strategy_dist}
\end{figure}

The \textit{Dependency Update Strategy} encapsulates the dependency constraints used by the npm package manager to determine acceptable versions of each package dependency \cite{jafari2023package, decan2019package}. These constraints specify the degree of freedom given to the package manager to automatically install new versions of a dependency and it is an important part of the dependency management practices in each project. We hypothesize that limiting automatic updates for dependencies increases the likelihood of depending on a vulnerable version of a dependency that has already released a fix. In the following, we elaborate on the specific definition of our three dependency update strategies. Figure~\ref{fig:strategy_dist} presents the distribution of dependency update strategies in our dataset.

\begin{itemize}
	\item \textbf{Balanced Update Strategy:} This update strategy encompasses dependency constraints that allow the package manager to automatically install new minor and patch releases for post-1.0.0 package dependencies and prevent automatic updates for pre-1.0.0 package dependencies (because SemVer considers pre-1.0.0 releases to have an unstable API \cite{semver2019}. A common practice to use a balanced update strategy in npm is to use the caret symbol as a dependency constraint (e.g. \textsuperscript{$\wedge$}1.2.3).
	\item \textbf{Restrictive Update Strategy:} This update strategy encompasses dependency constraints that either prevent automatic updates entirely or restrict the package manager to only install new patch updates for post-1.0.0 package dependencies. The tilde notation is commonly used in npm to specify restrictive update strategies (e.g. $\sim$1.2.3). We do not define restrictive strategies for pre-1.0.0 package dependencies as any automatic updates for such packages is considered permissive. 
	\item \textbf{Permissive Update Strategy:} This update strategy encompasses dependency constraints that allow the package manager to automatically install all new releases (including major releases) for post-1.0.0 package dependencies and any dependency constraint that allows updates of any kind for pre-1.0.0 package dependencies. The npm ecosystem allows the use of wildcards (e.g. *) as a dependency constraint but one can also use $>$= (e.g. $>$=1.2.3).
  \end{itemize}

In order to curate an accurate depiction of the relationship between features and vulnerability fixes, we need to calculate downstream package features at the time a vulnerability was fixed in an upstream package (e.g. we are interested in the release frequency of the downstream package at the time the upstream package released the fix, not many years after). Since the same npm package can depend on multiple vulnerable dependencies at different times, the features have to be calculated separately every time.

The dependency update strategy is a categorical feature with no ordered relationship and we must use one-hot encoding to encode each class into a binary (0,1) feature. Otherwise, the model may assume a natural order between the update strategies, which might result in poor performance or irrational results. We used the pandas library to retrieve one-hot encoded values for the update strategies \cite{pandas_onehot}. 


Introducing highly correlated features while training a machine learning model can impact both its performance and its interpretability. Due to the skewed distribution of our features, we use Spearman's correlation \cite{hollander2013nonparametric} to identify and remove features with a correlation score of above 0.7. 
In such cases, we kept the feature which we believed to have a more tangible definition. We dropped \textit{Package Version Count} in favor of \textit{Dependency Modifications} and \textit{Days Since Last Release} in favor of \textit{Package Age}. The final set of features for our study is presented in Table~\ref{tab:selected_features}. It is worth noting that some of the collected features represent the characteristics of a selected dependency (e.g. Age and Dependent Count) while others reflect their behavior (e.g. Update Strategy and Release Frequency).


\begin{table*}
	\footnotesize
	\centering
	\caption{Selected features for downstream packages}
	\label{tab:selected_features}
	\begin{tabular}{llc}
		\toprule
		Feature & Description & Histogram\\
		\midrule
		Package Age & The number of days since the dependent package was published. & \includegraphics[width=0.1\linewidth]{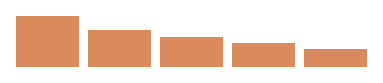}\\
		Balanced Update Strategy & Whether or not the dependent used the balanced update strategy. & \includegraphics[width=0.1\linewidth]{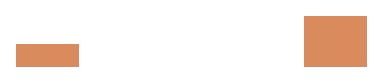}\\
		Restrictive Update Strategy & Whether or not the dependent used the restrictive update strategy. & \includegraphics[width=0.1\linewidth]{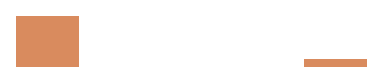}\\
		Permissive Update Strategy & Whether or not the dependent used the permissive update strategy. & \includegraphics[width=0.1\linewidth]{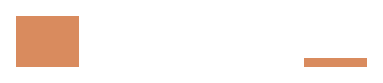}\\
		Release Frequency & The average number of dependent version releases per month. & \includegraphics[width=0.1\linewidth]{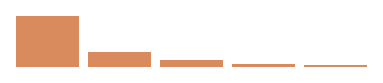}\\
		Dependency Count & The total number of dependencies for the dependent package. & \includegraphics[width=0.1\linewidth]{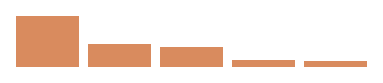}\\
		Dependent Count & The total number of dependents for the dependent package. & \includegraphics[width=0.1\linewidth]{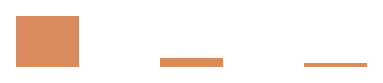}\\
		Release Status & Whether or not the dependent package was in a pre-1.0.0 or post-1.0.0 release state. & \includegraphics[width=0.1\linewidth]{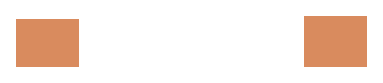}\\
		Dependency Modifications & The number of dependent releases in which the dependency configuration was modified. & \includegraphics[width=0.1\linewidth]{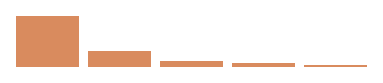}\\
		\bottomrule
	\end{tabular}
\end{table*}

\section{Results}
\label{sec:Results}
In this section, we motivate our 3 research questions, describe our approach, and present our findings.
\subsection{\rqi}
\label{sub:rq1}

\noindent\textbf{Motivation:}
Security vulnerabilities rooted in dependency packages are a major risk to software \cite{zimmermann2019small, sonatype2021}. Packages play a key role in mitigating vulnerabilities in their dependencies. We want to understand the role of dependency management decisions of packages in their exposure to vulnerabilities from upstream packages.
While previous research has discussed how long it takes for a npm dependency to remain vulnerable \cite{alfadel2023discoverability, decan2018impact, chinthanet2021lags}, they have not studied the relationship between vulnerabilities and the dependency management decisions of downstream dependents.

\begin{figure*}
	\centering
		\begin{subfigure}{.23\linewidth}
			\includegraphics[width=\linewidth]{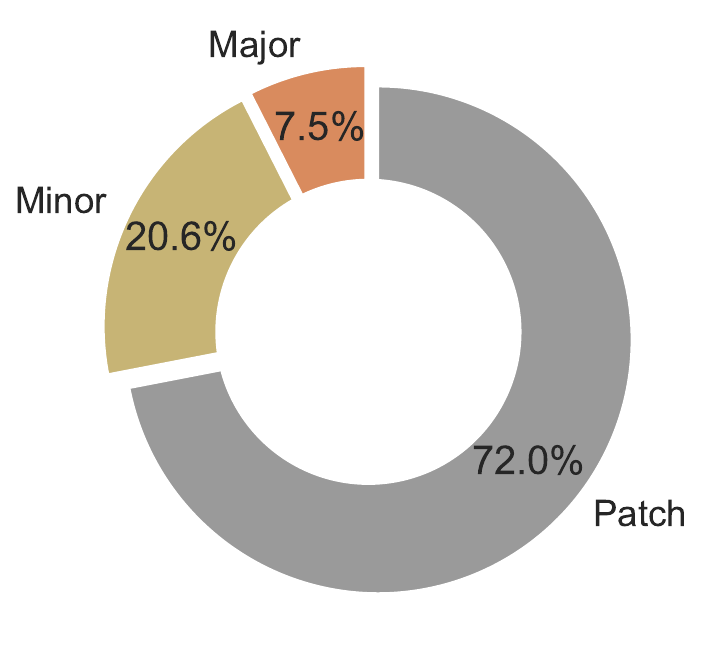}
			\caption{Critical}
		\end{subfigure}
		\begin{subfigure}{.23\linewidth}
			\includegraphics[width=\linewidth]{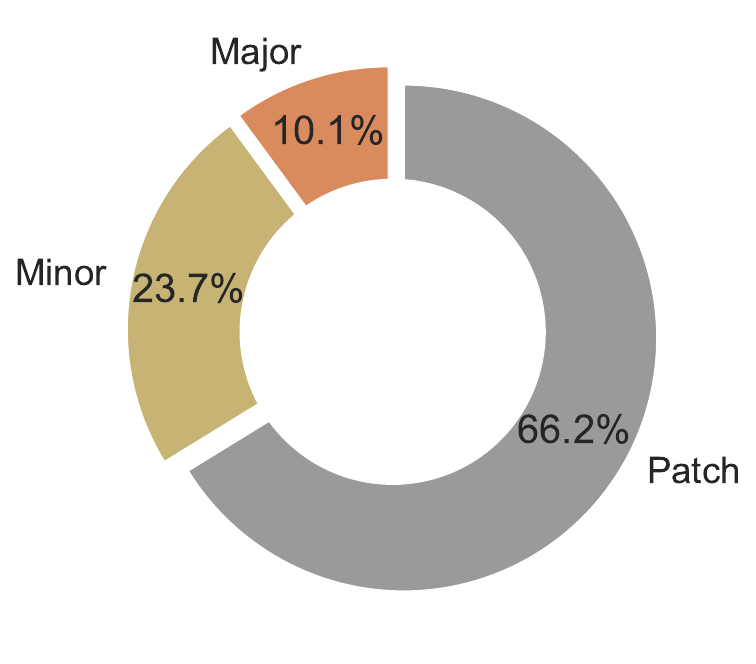}
			\caption{High}
		\end{subfigure}
		\begin{subfigure}{.23\linewidth}
			\includegraphics[width=\linewidth]{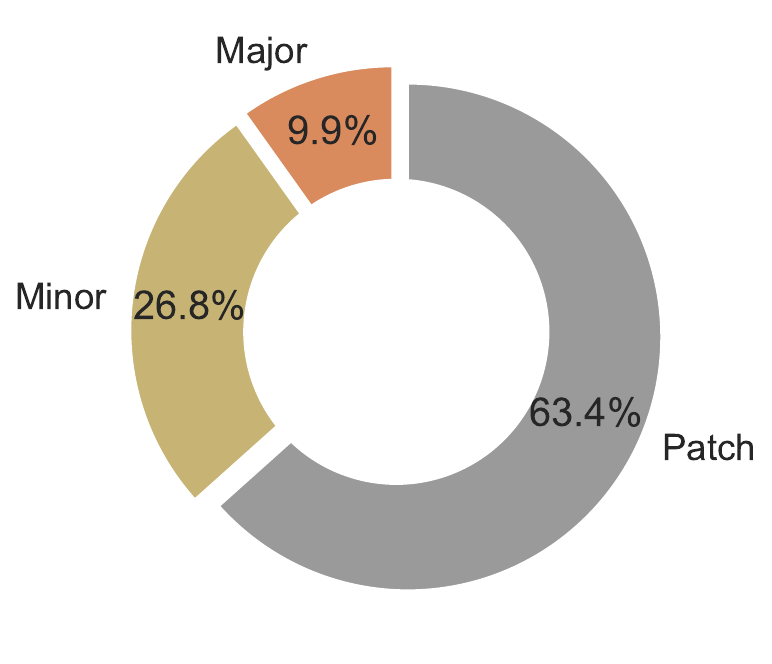}
			\caption{Medium}
		\end{subfigure}
		\begin{subfigure}{.23\linewidth}
			\includegraphics[width=\linewidth]{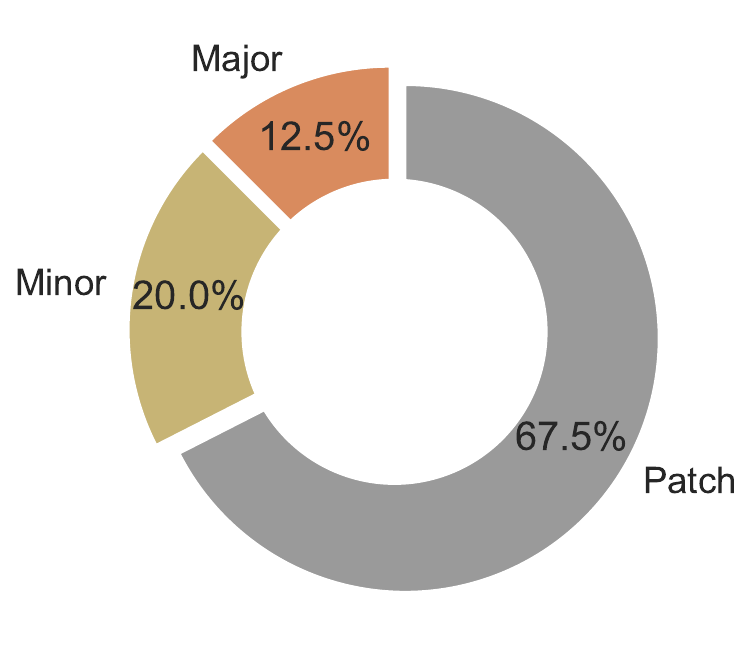}
			\caption{Low}
		\end{subfigure}
	
	\caption{Release type for the vulnerability fix (separated by vulnerability severity)} 
	\label{fig:release_dist}
\end{figure*}

\noindent\textbf{Approach:}
A vulnerability fix can be released as a major, minor or patch version (Section~\ref{sec:Background}). Since we have the information for package versions and vulnerability metadata, we can compare the version number of the fixing release (r) with the version number of the release right before the fix (r-1) to evaluate the type of fixing release. 
(Figure~\ref{fig:release_dist}). We use the npm SemVer package \cite{pkg_semver} to evaluate the difference between these two packages. For this analysis, we exclude uncommon release types such as pre-major and pre-minor and focus on the main types of releases according to semantic versioning guidelines \cite{semver2019} (i.e. major, minor and patch). Since the adoption of different release types is determined by the dependency decisions of downstream dependents, comparing the proportion of release types for vulnerability fixes shows how downstream dependency management decisions can influence the proportion of vulnerability mitigation. 

To evaluate the speed of vulnerability mitigation, we measure how long it takes for the upstream packages to fix the vulnerability after public disclosure \textbf{(i.e. fix delay)} and how long it takes for the downstream dependents to adopt the fix after the fix is released \textbf{(i.e. adoption delay)}. Since the adoption of a fix is the responsibility of downstream dependents (through dependency management decisions), comparing the fix delay and adoption delay shows how much downstream dependency management decisions influence the total mitigation time.

The fix delay is measured by comparing the public exposure date and the fix release date which are both provided by the vulnerability advisory.
In order to measure the adoption delay, we determine which dependents actually install a vulnerable version of the upstream vulnerable package and later adopt the fix (Section~\ref{sec:data-curation}). We then look at the release date of the fix and compare it with the first release of the dependent that accepts the upstream fix.
In addition, 74,909 dependent packages in our dataset never adopt a fix, so we measure the adoption delay from when the fix was published to the most recent date in the dataset (12 Jan 2020). This is a lower bound for the adoption delay as the dependents can opt to receive a fix at a later date.

\begin{figure*}
    \centering
    \includegraphics[width=0.96\linewidth]{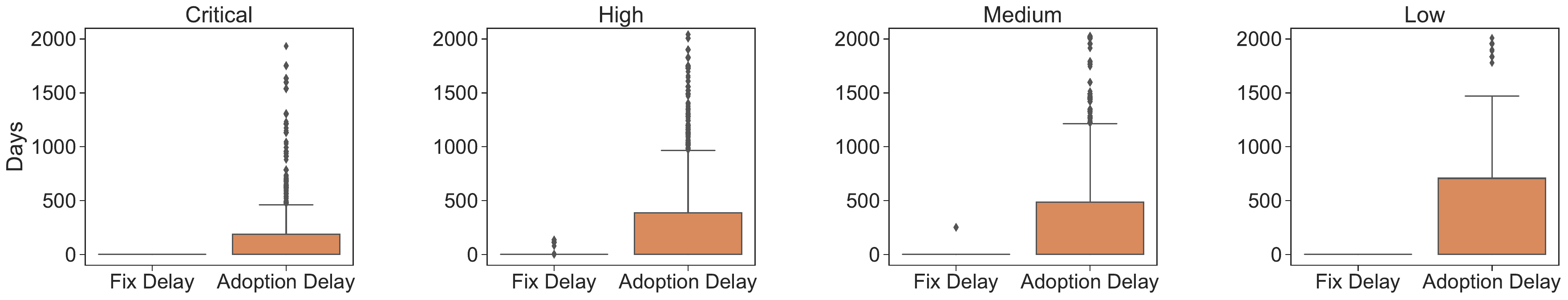}
    \caption{Comparing the distributions of upstream fix delay with downstream adoption delay for different vulnerability severities (statistically significant difference with p$<$0.05). }
    \label{fig:compare_delay}
\end{figure*}

Figure~\ref{fig:compare_delay} presents the distributions for the fix and adoption delays for different vulnerability severities. We use the Mann-Whitney U test to measure the statistical significance of the difference between the distributions \cite{mann1947test}. The distributions should not be biased towards popular vulnerabilities that impact many dependents or the decisions of packages that have many vulnerable dependencies. In order to prevent a bias towards vulnerabilities that affect many dependents, the distributions of fix delays are calculated for \textbf{unique vulnerabilities}. Similarly, to prevent a bias towards downstream packages that are affected by many vulnerabilities, the distributions of adoption delays are calculated for \textbf{unique dependent packages} (i.e. when a package has multiple vulnerable dependencies, we consider the most recent vulnerable dependency).

\noindent\textbf{Findings:}
The release types of the vulnerability fixes for each severity class are depicted in Figure~\ref{fig:release_dist}. The majority of vulnerability fixes are released as a patch update, which is expected as these types of releases are meant for fixing bugs and vulnerabilities. Yet, a sizeable portion (more than 30\% on average) of vulnerability fixes are released in minor and major release types, which aligns with the findings of previous research \cite{chinthanet2021lags, zerouali2022impact}. This creates a problem for downstream dependents because previous research has shown that even though dependents frequently use the non-latest version groups, back-porting fixes to previous version groups is an infrequent practice \cite{decan2021back}. Consequently, downstream dependents that refuse to accept all updates will not receive all vulnerability fixes. In other words, downstream dependency practices impact what proportion of vulnerability fixes can be adopted. The solution is not as straightforward as ``accepting all updates'', as new \textit{major} versions will (by definition) contain backward incompatible changes that can break the downstream package. Even new minor or patch releases may introduce such breaking changes \cite{mezzetti2018type,cogo2019empirical}. 


\conclusion{Finding 1: Since vulnerability fixes are packaged in different release types, downstream dependency management decisions that limit updates to specific release types (e.g. patches) also limit their capacity of receiving the fixes.}

The more important finding is the relationship between downstream dependency management decisions and the adoption delay of vulnerability fixes. The box plots in Figure~\ref{fig:compare_delay} provide insights into how long npm packages remain vulnerable. 
For each severity type, we have plotted the distribution for the fix delay and the adoption delay. While the median for all delays is zero (meaning many publicly disclosed vulnerabilities are fixed within one day and adopted by the dependents on the same day), the distribution of the delays clearly show that adoption delay is usually much longer than the fix delay. The difference between two distributions are statistically significant (for all severities) using the Mann-Whitney U test with p$<$0.05. The mean time to release a fix is 1.17 days, while the mean time to adopt a fix is 192.5 days. While only 4 vulnerabilities take more than a day to release a fix (after disclosure), 73,155 dependent packages need more than a day to adopt an upstream fix. 

The minuscule delay to release a fix is due to the vulnerability disclosure policies that recommend the initial vulnerability report to be made privately to package maintainers so they can release a fix before publicly disclosing the vulnerability \cite{vul_policy}. In fact, apart from a few outliers, almost all publicly disclosed vulnerabilities have a vulnerability-fixing release on the same day. This is not the case for the adoption of vulnerability fixes. It is the delay in downstream fix adoption, not the delay in the upstream fix release, that is keeping dependent packages vulnerable. What makes matters worse, is that the post-disclosure vulnerability risk period is more dangerous than the pre-disclosure period because the details of the vulnerability and the means to exploit it are now made public. A crucial factor in determining how fast publicly disclosed vulnerabilities are mitigated across the ecosystem is for downstream dependents to adopt appropriate dependency management practices.

One interesting observation in Figure~\ref{fig:compare_delay} is that the delay in fix adoption corresponds intuitively and consistently to the severity of the vulnerability. The lower the severity of a vulnerability, the higher the delay in adopting the fix. This hints that developers are aware of vulnerability severities and use the information to prioritize the adoption of fixes. As previous research has shown, developers often do not update their dependencies due to the necessary extra effort \cite{kula2018developers}. The built-in \textit{npm audit} command scans package dependencies for known vulnerabilities and reports the list and severity of public vulnerabilities \cite{npm_audit}. A similar functionality is provided by the popular Dependabot dependency management tool \cite{dependabot}.


\conclusion{Finding 2: Since the majority of packages with a vulnerability release a fix within a day of public disclosure, downstream dependency management decisions that delay fix adoption are the key culprit for prolonged exposure to such vulnerabilities.}

\subsection{\rqii}





\noindent\textbf{Motivation:}
In our first research question, we find that dependency decisions impact the responsiveness of packages to adopting a vulnerability fix. We need to identify the attributes of responsive dependency packages. By identifying responsive packages, we can help developers mitigate the risk of vulnerable dependencies. We wish to use our selected package attributes to model the adoption delay of vulnerability fixes. Our results will help developers better understand how opting for different dependency practices (e.g. choosing a package with a higher release frequency) can increase or decrease the adoption speed of vulnerability fix propagation from upstream packages.

\noindent\textbf{Approach:}
We use our set of 9 package attributes (Section~\ref{sec:data-curation}) to train a random forest model to predict the adoption delay of the fix. We use random forest because it is known to offer a good balance between performance and interpretability and is commonly used in software engineering research \cite{ohm2022feasibility, dey2020deriving}. When feeding the vulnerability instances to the model, we use the latest vulnerable dependency relationship for each downstream dependent. We do this as we do not want non-unique downstream packages across our training and test data and we do not want to bias the data towards downstream packages with a high number of dependencies. For example, package X may depend on a vulnerable package in 2016 and another vulnerable package in 2019. Since each dependency relationship is an instance used to train the model, relationships involving package X may appear in both training and test sets. While the features for package X are calculated separately per dependency relationship, a hidden feature (e.g. cultural habits) may remain consistent for both relationships, causing a data leak from the training set to the test set. Additionally, package X may have many more vulnerable dependencies than another downstream package, which would bias the model results to the features of package X due to increased presence in the training set.
This amounts to 201,027 vulnerable dependency relationships in our data. Since no previous work has used dependency practices to model adoption delay, we use a stratified predictor (random prediction based on class weight) as our baseline \cite{scikit_dummy}. Both our random forest and our baseline model are trained and tuned on 80\% of our dataset (training and validation set) and evaluated on the held-out 20\% (test set) \cite{geron2019hands}. We use ROC-AUC and F1-score to evaluate our models \cite{tharwat2020classification}. The ROC metric (Receiver Operating Characteristics) depicts a probability curve and the AUC (Area Under the Curve) is a value in the range of 0 and 1 that shows the capability of the model in distinguishing between classes. Higher ROC-AUC means the model is better at correctly predicting classes. F1-score is a function in the range of 0 and 1 that measures the balance between precision (the portion of true positive cases among all the retrieved cases) and recall (the portion of true positive cases that were retrieved). 

We used a grid search with a 10-fold cross validation on the training set to tune the hyper-parameters of our model. This results in 1000 estimators (trees) with a minimum sample split of 8. The 10-fold cross validation fits the model 10 times, where each fit is performed on 90\% of the training set (randomly selected) and the remaining 10\% is used as a validation set.

\begin{figure}[h]
    \centering
    \includegraphics[width=\linewidth]{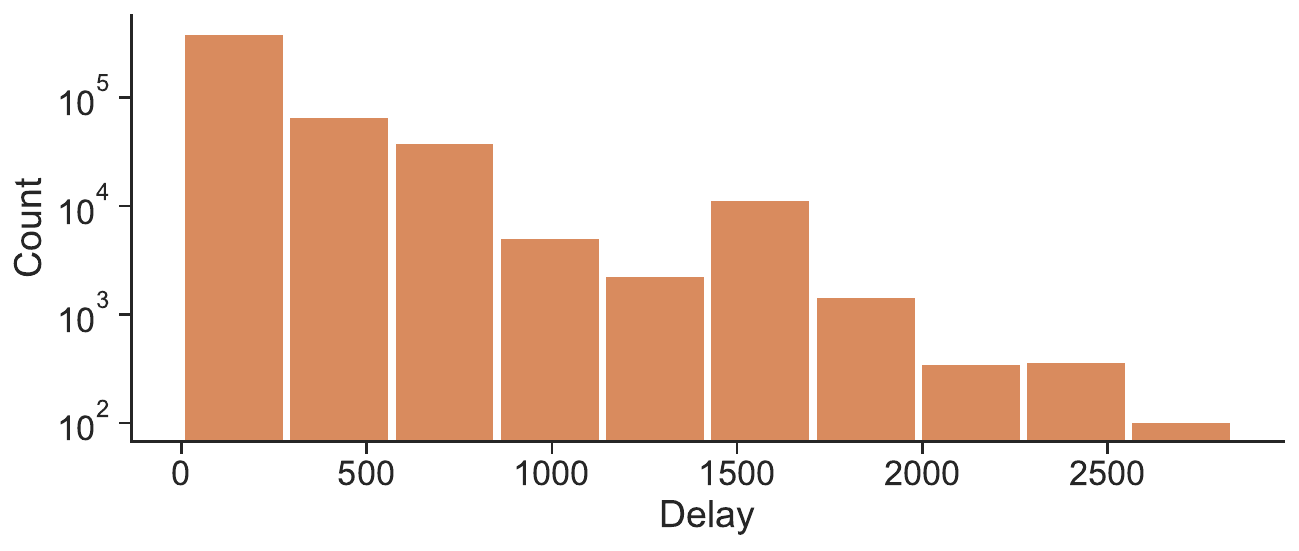}
    \caption{Distribution of adoption delay (days)}
    \label{fig:delay_dist}
\end{figure}

We are interested in predicting a range of classes of adoption delay (e.g. fast, slow), instead of the actual number of days because the adoption delay ranges from 0 to 2841 days and the delay is not equally distributed along the range.
Figure~\ref{fig:delay_dist} presents the distribution of delay in 10 bins. As can be seen, the  majority of vulnerability fixes are adopted in a short time (note the log scale on the Y-axis of Figure~\ref{fig:delay_dist}). 

We train a binary classifier and use a threshold of 48 hours for identifying the fast-responder class. Since it is not intuitive to classify anything below 48 hours as fast and anything slightly above it (e.g. 50 hours) as slow, we use a disjoint threshold of below 48 hours for the fast class and above 14 days for the slow class to better separate the classes. This removes 546 dependents between the two thresholds (that adopted the fix later than 48 hours and sooner than 14 days) from our dataset. We conducted a sensitivity analysis to analyze our binary class threshold and ensure minor changes do not significantly impact the target class distribution (see Section~\ref{sec:Threats to Validity}).

\begin{figure}[h]
    \centering
    \includegraphics[width=\linewidth]{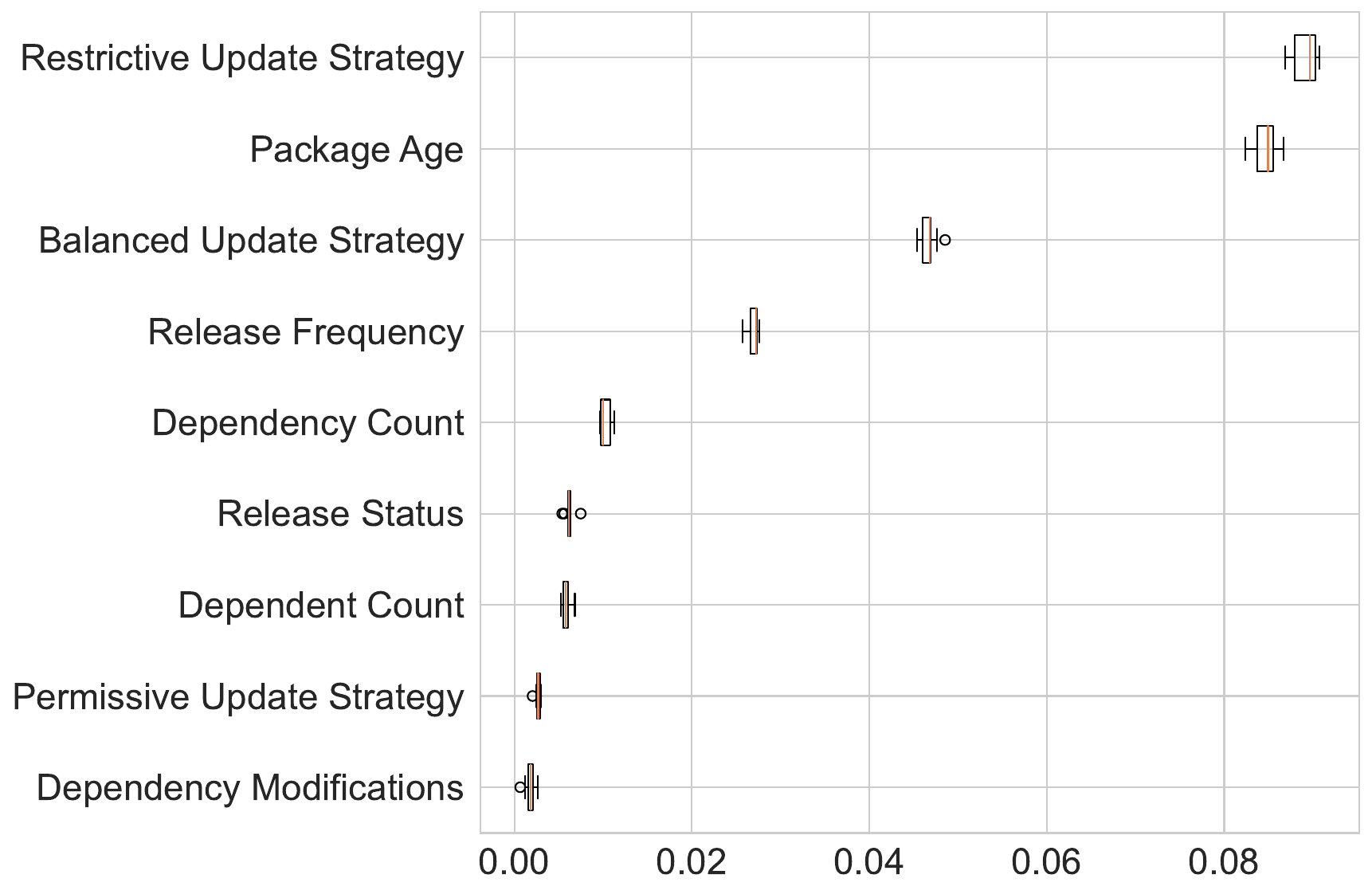}
    \caption{Ranking the features of the model based on permutation importance}
    \label{fig:feature_importance}
\end{figure}

Since not all dependency practices have the same impact in predicting the adoption delay, we need to identify, rank and analyze the important features of the model (Figure~\ref{fig:feature_importance}). We calculate the permutation feature importance on the test set in which each feature is randomly shuffled (repeated 10 times) to observe its impact on the model's performance (ROC-AUC). Important features have a larger influence on the model's performance when their values are permuted \cite{featureimportance}. 



\begin{figure*}[h]
    \centering
    \includegraphics[width=\linewidth]{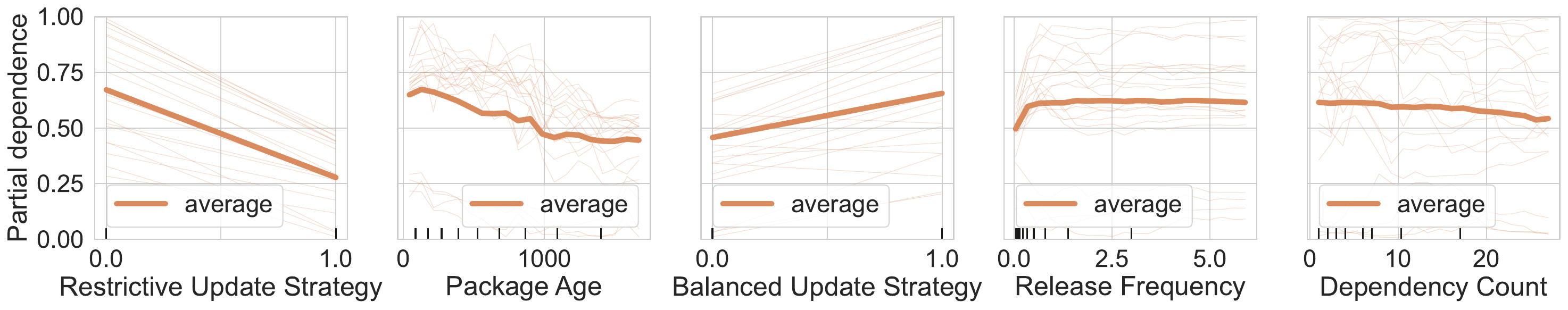}
    \caption{Partial Dependence Plots and Individual Conditional Expectations for the top 5 dependency practices}
    \label{fig:pdp}
\end{figure*}

We use Partial Dependence Plots (PDP) to visualize the impact of the important dependency practices on the predicted adoption delay (Figure~\ref{fig:pdp}). PDPs depict the marginal effect of a feature on the model's predictions \cite{molnar2020interpretable} and they can highlight linear, monotone or more complex relationships between the dependency practices and the adoption delay. In other words, we can visualize how a change in a feature can change the adoption delay predicted by the model. The Y-axis on the PDPs in Figure~\ref{fig:pdp} is the predicted probability for an instance being predicted as a fast response. The tick marks on the X-axis are the deciles for the feature distribution which indicate which part of the plots represent the majority of our dataset. Individual Conditional Expectation (ICE) plots also show the effect of a feature on the target variable. However, unlike PDP which average the effect over all instances, ICE plots visualize the relationship for a single instance \cite{goldstein2015peeking}. We have shown a random sample of 20 ICE plots in Figure~\ref{fig:pdp} for each of the top 5 features (depicted using the light-colored lines).


\begin{figure}[h]
    \centering
    \includegraphics[width=0.73\linewidth]{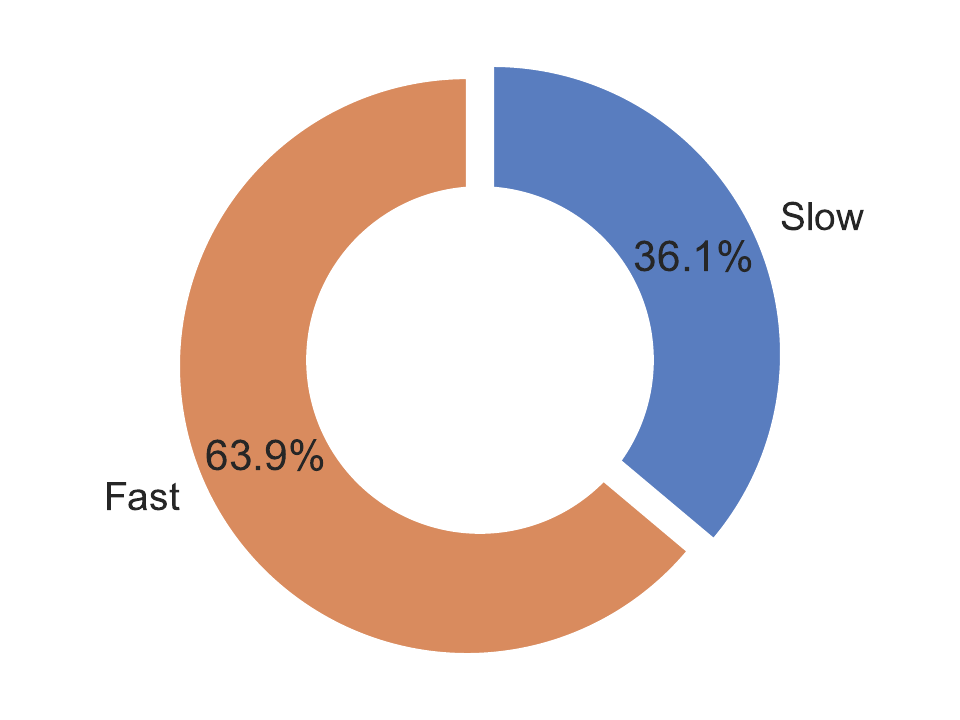}
    \caption{Distribution of model classes}
    \label{fig:class_dist}
\end{figure}

\begin{figure}[h]
    \centering
    \includegraphics[width=\linewidth]{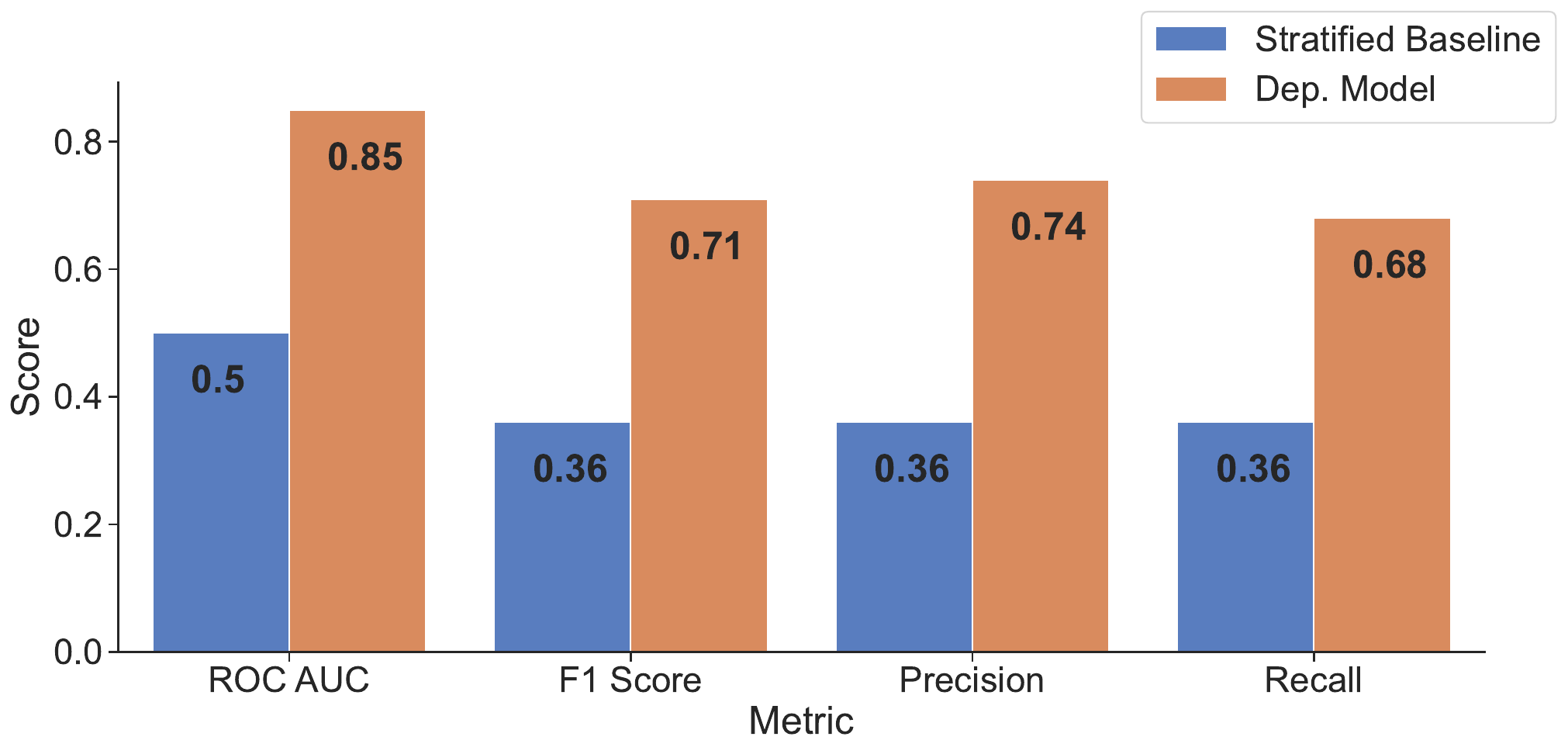}
    \caption{Performance evaluation results for the dependency practices model}
    \label{fig:eval_all}
\end{figure}

\noindent\textbf{Findings:}
The evaluation results in Figure~\ref{fig:eval_all} present the ROC-AUC, F1 score, Precision and Recall for our dependency practices model compared with the stratified baseline. Our dependency practices model achieves an ROC-AUC of 0.85, which is a 70\% improvement over the baseline. Our model also achieves an F1-score of 0.71, which is a 97\% improvement compared to the baseline. These results indicate that package features can be used to model and predict the adoption delay of vulnerability fixes in the npm ecosystem. 


We also train and evaluate the model on four subsets of the data in Table~\ref{tab:dataset_stats}, which separates the dataset based on vulnerability severity to see if there is a considerable difference between our results when we focus on critical, high, medium and low severity vulnerabilities. As can be seen in the evaluation breakdown of Table~\ref{tab:eval_breakdown}, our data subset models perform at least on par with the main model. In some cases, we observe even stronger performance results (e.g. ROC-AUC of 0.92 for critical vulnerabilities).

\begin{table*}[t]
	\centering
	\caption{Performance evaluation of alternative models on severity-specific subsets of the data}
	\label{tab:eval_breakdown}
	\begin{tabular}{l|r r r|r r r}
	\toprule
	\centering
	\textbf{Subset model} & \textbf{Model ROC-AUC} & \textbf{Baseline ROC-AUC} & \textbf{Improvement} & \textbf{Model F1-score} & \textbf{Baseline F1-score} & \textbf{Improvement}\\
	\midrule
	
	Critical & 0.92 & 0.5 & 84\% & 0.82 & 0.35 & 134\% \\
	High & 0.85 & 0.5 & 70\% & 0.73 & 0.38 & 92\% \\
	Medium & 0.91 & 0.5 & 82\% & 0.80 & 0.34 & 135\% \\
	Low & 0.92 & 0.5 & 84\% & 0.83 & 0.41 & 102\% \\
	\bottomrule
\end{tabular}

\end{table*}

\conclusion{Finding 1: Practitioners can use the attributes of their dependency packages to influence the adoption of upstream vulnerability fixes.}

We have ranked the dependency practices based on their permutation importance in the box plots of Figure~\ref{fig:feature_importance}. 
The use of a restrictive update strategy, the age of the package, the use of a balanced update strategy, the release frequency of the package and the number of dependencies for a package are the most important indicators for determining the responsiveness of an npm package to a vulnerability fix. Developers should incorporate these characteristics in their dependency management and selection practices.

While feature importance tells us which dependency practices are instrumental for the model, the PDPs in Figure~\ref{fig:pdp} tell us how the dependency practices impact the prediction results. Specifically, we want to analyze how a change in the top 5 dependency practices increase or decrease the likelihood of a package being classified as a fast responder to vulnerabilities. We can see that packages with restrictive update strategies take longer to adopt vulnerability fixes. Consequently, their increased exposure to vulnerabilities further exposes their downstream dependents (transitive exposure). While this observation aligns with our intuitive understanding of restrictive updates, it is not always the expected outcome. Developers that use restrictive update strategies can always decide to manually maintain their dependencies and keep track of important updates, but in a generalized sense, they fall short of the latest vulnerability fixes; especially compared to developers who opt for a more balanced update strategy. 

\conclusion{Finding 2: Downstream dependents of packages with a non-restrictive dependency update strategy tend to have shorter exposure to security vulnerabilities.}

A less intuitive finding is the impact of package age on the responsiveness to vulnerabilities. Older packages in our dataset are predicted less likely to quickly adopt vulnerability fixes. We believe this is due to the impact of old and unmaintained packages, since a well-maintained mature package (while seemingly rare) can actually be a more reliable dependency choice. This hypothesis is reinforced by the PDPs for release frequency. Packages that release more often are more likely to be predicted as a fast responder. It is also worth noting the initial feature of ``Days Since Last Release'' was removed due to a high correlation with ``Package Age'', indicating that older packages are more likely to not have a recent release.

\conclusion{Finding 3: Downstream dependents of younger packages tend to have shorter exposure to security vulnerabilities.}

A higher release frequency of a package initially increases the likelihood of a fast response to vulnerabilities. However, once the package has a new release every 4 months, a further increase in release frequency does not seem to make a difference. There are two potential explanations for this observation. First, a non-zero release frequency of around 0.25 \textit{per month} is already enough to differentiate between regularly maintained and abandoned packages and further increase in the release frequency may not provide further insight into adequate maintenance. The second explanation for this observation is the tick marks on the X-axis, which represent the distribution deciles for the feature. As can be seen, the majority of the distribution is collected in the left side of the PDP, indicating that the initial positive slope is in fact a better representation of the overall effect of release frequency on the likelihood of a fast response.

\conclusion{Finding 4: Downstream dependents of packages with a faster release cycle tend to have shorter exposure to security vulnerabilities.}

Packages with a higher number of dependencies are less likely to be predicted as a fast responder. Having more dependencies can make it more difficult to keep track of (and respond to) vulnerable dependencies in a timely manner, especially if relying on manual updates. The PDP slope in Figure~\ref{fig:pdp} only indicates a modest effect by dependency count that does not influence the model's predictions as strongly as the other top features. It is also worth noting that our findings are focused on the delay of adopting vulnerability fixes, but having more dependencies can also increase the threat surface for vulnerable dependencies.

\conclusion{Finding 5: Downstream dependents of packages with less dependencies tend to have shorter exposure to security vulnerabilities.}

\subsection{\rqiii}
\label{section:rqiii}

\noindent\textbf{Motivation:}
The key objective of this study is to assist developers in identifying responsive packages to reduce the risk of vulnerability exposure from dependencies. 
Therefore, we need to understand the perception of developers regarding our findings. Are developers aware of the impact of downstream dependency practices on vulnerability mitigation? Do our findings reinforce or contradict their experiences? Will they use our findings to complement their dependency selection process? Answering these questions will help us understand the applicability of our findings in practice. 

\noindent\textbf{Approach:}
To understand the perception of developers on the use of dependency management practices for mitigating vulnerabilities, we crafted a survey that aims to compare the findings of RQ1 and RQ2 with the real-world dependency management experience of practitioners. 

The survey is composed of 4 parts. In the first part, we ask respondents about their background and experience. In part 2, we ask the participants about their opinion (based on their experience) on the relative effect of publishing the fix compared to adopting the fix as the main reason for exposure to publicly disclosed vulnerabilities (RQ1). In part 3, we ask the participants about the impact of the features used for our model (RQ2) on the responsiveness to vulnerability fixes. In the final part, we provide respondents with the initial findings of our study and ask if they would incorporate our findings in their dependency selection and management practices in the future. 

All of the questions included an "other" option which allowed developers to provide an answer not already included in the choices or to expand on their answer. We also asked respondents if they wish to suggest additional package attributes that affect the response speed to vulnerable dependencies. Some of these free-form responses are exclusively referenced in Section~\ref{sec:Implications}. The complete set of questions and response choices are included in our replication package \cite{replication_vul}.

In order to recruit the participants for the survey, the authors distributed it among their existing network of industry developers and academic researchers (mostly in Canada) which are actively working in the industry or academia. We contacted 114 practitioners and received a total of 67 responses.
As Table~\ref{tab:survey_background} shows, the respondents are primarily composed of industry practitioners (59.7\%), followed by academic researchers (29.8\%). The majority of our participants (67.2\%) have 4 years or more experience in software development.

\begin{table}
	\centering
	\caption{Background of participants in the survey.}
	\label{tab:survey_background}
	\begin{tabular}{p{1.4in}p{1.3in}r}
	
	\toprule
	\textbf{Dimension}	& \textbf{Responses} &  \textbf{\%}\\
	
	\midrule
	\multirow{3}{*}{Background}		
	& Industry Practitioner			&		59.7\% 	\\
	& Academic Researcher			&		29.8\% 	\\
	& Both							&		7.5\% 	\\
	& Student						&		3\% 	\\

	\midrule 
	\multirow{4}{*}{Development Experience}		
	&	$>=$ 7 years				&		38.8\% 	\\
	&	4-6 years					&		28.4\% 	\\
	&	1-3 years					&		31.3\% 	\\
	&	$<$ 1 year					&		1.5\%	\\
	
	\midrule
		\multicolumn{2}{l}{\textbf{Total participants}}  & \textbf{67} \\
	
	\bottomrule
\end{tabular}
\color{black}

\end{table}

\noindent\textbf{Findings:}
In the following, we will present the findings of our practitioner survey. We have also included sample responses from the participants. We empirically discovered (RQ1) that it is the adoption delay, not the fix delay, that is the main contributing factor for the survival of publicly disclosed vulnerabilities in the npm dependency ecosystem. This aligns with the experience of a considerable portion (46.3\%) of our respondents. However, 40.3\% of the respondents believe that both the fix delay and adoption delay are equally responsible for exposure to publicly disclosed vulnerabilities. Surprisingly, 10.4\% of respondents believe the fix delay to be the main contributing factor. Respondents also highlighted the criticality of the downstream dependent in the outcome. 

\highlight{``The projects deployed in industry are usually relatively large and extensive. [...] nobody dares to manipulate dependencies as any change may stop the system's regular functionality'' - R12}

Indeed, updating dependencies is a more sensitive decision when backward compatibility is crucial.

\highlight{``It depends on the project. For some projects, the security level is critical, so if there is any known vulnerability, it can be more option one [fix delay], and for less critical project, more option 2 [adoption delay].'' - R58}

In other words, a security critical downstream client is less likely to be exposed to a vulnerability through their own fault, but rather because the fix is not yet released. 

\conclusion{Finding 1: The majority of the practitioners do not believe that the delay in downstream ﬁx adoption is more responsible than the delay in the upstream ﬁx release in keeping dependent packages vulnerable.}

The next section of the survey focuses on the important package features for predicting the adoption delay (RQ2). The responses align with our findings for RQ2, both for the importance of the features and for the positive/negative relationship between the features and the response delay. When asked about the relationship between package age and the speed of handling vulnerable dependencies, 46.3\% of practitioners believed older packages to be slower in addressing vulnerable dependencies, which aligns with our finding in RQ2. Only 10.4\% believed older packages to be faster in addressing vulnerable dependencies. 37.3\% of the practitioners selected the ``Neither'' option. Respondents also highlighted the distinction between old active packages and old unmaintained packages. 

\highlight{``[...] I believe packages that see active development are more likely to address vulnerable dependencies'' - R20}

\highlight{``If its really old then maybe its kinda abandoned but if its too young then it might be immature'' - R39}

When asked about the relationship between release frequency and how fast vulnerable dependencies are addressed, 85.1\% of practitioners believed packages that release more often are faster in handling vulnerable dependencies. Respondents also highlighted the importance of the reason behind a frequent release cycle.

\highlight{``packages in early development release more often. Not to fix vulnerabilities but to change features. After that initial period release frequency might be more related to vulnerabilities'' - R66}

\highlight{``[...] a healthy release cycle is good but too much may just be a sign of bad versioning practice by the devs'' - R39}

Our findings in RQ2 show that release frequency does have a positive relationship with a fast response to vulnerabilities, but only to an extent. We observed no further improvement after a release frequency of around 0.25 per month. 11.9\% of practitioners selected the ``Neither'' option.

We previously found in RQ2 that higher dependency count in our model decreases the likelihood of a fast response to vulnerabilities. In the survey, 64.2\% of the practitioners also believe packages with a high number of dependencies are slower in addressing vulnerabilities. 19.4\% believe the opposite is true and 11.9\% believe there is no difference either way. Practitioners highlighted that the number of dependencies should be viewed in conjunction with project size and policies.  

\highlight{``If the dev team is small then more dependencies will make it harder to keep track so it might increase risk but a large dev team should maintain good dependency health anyway'' - R66} 

\highlight{``Depends on project/org policies'' - R16} 

This may explain the reason we observed a \textit{weak} relationship between dependency count and the response to vulnerabilities in RQ2.

We asked practitioners to rank the various dependency update strategies based on which update strategies in a package lead to faster handling of vulnerable dependencies with a publicly disclosed vulnerability. Among the respondents, 58.2\% agree that the balanced update strategy leads to a faster response to vulnerable dependencies whereas a restrictive update strategy leads to a slower handling of such vulnerable dependencies. Going further than a balanced update strategy and adopting a more permissive strategy allows for even more updates but increases the risk of breaking changes. 

\highlight{``The best practice is to allow automatic updates for new patch and minor versions [i.e. balanced strategy]. For major changes, it is not technically possible to allow automatic updates since the major updates [i.e. permissive strategy] include breaking changes for other dependencies [...]'' - R38}

Indeed, developers are aware that fixes are not always released in (or backported to) patch releases, but they are wary of the trade-offs (between receiving all fixes and breaking changes) for allowing too much freedom in automatic updates. 

\highlight{``believe it or not, some fixes are in major updates! but its usually not a good idea to update all the time like that'' - R39}

Even though we consider both patch only updates and no updates as restrictive update strategies, we gave respondents the option to rank these two approaches separately. 86.6\% of respondents ranked patch only updates as a better approach than no automatic updates.

\conclusion{Finding 2: The majority of practitioners agree that younger packages with frequent releases that adopt a balanced update strategy and have fewer dependencies are faster in addressing vulnerable dependencies.} 

Even though the Update Strategy, Package Age, Release Frequency and Dependency Count are the important features for our model (Figure~\ref{fig:feature_importance}), we asked practitioners about all the features in our study. When asked about the number of modifications to dependency configuration, 50.7\% of practitioners believe more frequent modifications of the dependency configuration file is associated with a faster response to vulnerable dependencies, while 32.8\% do not believe this feature to be relevant to assess a packages response to vulnerable dependencies. 

\highlight{``Depends. touching things too much could be a sign of diligence or an inexperienced dev'' - R66}

For dependent count, 44.8\% of respondents believe packages with higher dependent counts to be faster in addressing vulnerable dependencies whereas 26.9\% believe they would be slower to react. Having a large community can motivate the package maintainers to be more diligent since vulnerabilities from their dependencies can propagate to their large client base.

\highlight{``Usually yes because they get bombarded from their community if they don't [...]'' - R66}

In regards to release status, 37.3\% believe post-1.0.0 packages to be faster in addressing vulnerable dependencies while 38.8\% do not believe this feature to be relevant to how fast packages respond to vulnerable dependencies.

\highlight{``Some packages just enjoy staying in pre-1.0.0 (god knows why!) but they have a large following so they are good at handling vulnerable dependencies on time. Generally though, post-1.0.0 is better'' - R66}

\begin{figure}[h]
    \centering
    \includegraphics[width=\linewidth]{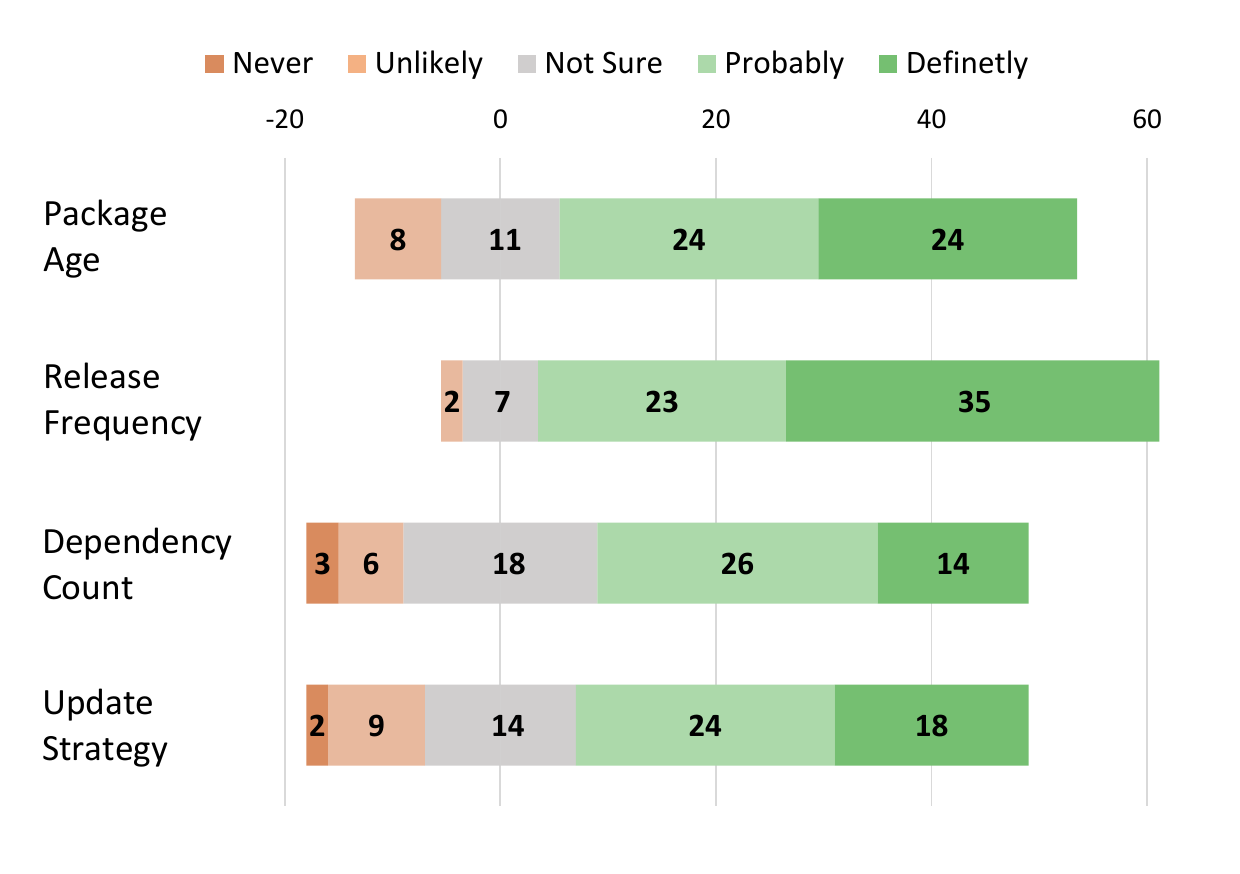}
    \caption{Likelihood of our top features being used in practice.}

    \label{fig:survey_agreement}
\end{figure}

In the final section of the survey (after asking the respondents about their perceptions on our proposed attributes) we presented our own findings on the most important package attributes that lead to better responsiveness to vulnerable dependencies. We asked participants how likely they are to use our proposed attributes in their dependency practices in the future. Figure~\ref{fig:survey_agreement} presents the results for each of attributes, ranging from Never using the attribute to Definitely using the attribute in the future. As can be seen, participants have a generally positive outlook on the applicability of our findings in practice. Specifically, practitioners have a strong tendency to use release frequency as a criterion for selecting packages in order to mitigate vulnerabilities from transitive dependencies.

While the top attributes in our model are a considerable indicator of how fast packages address vulnerable dependencies, they are not equally accessible to all developers. 

\highlight{``The age and release frequency are attributes that are often easy to check on a package management website or GitHub. These attributes are often good indicators of the health of the package, thus, the likelihood that maintainers will address vulnerable dependencies in a fast manner[...]'' - R22}

\highlight{``[...] Regarding dependency counts, I know having more dependencies to manage carries more risks, but sometimes there are specific packages you need that don't have other equivalents in npm. [...] I'm not sure if we can see somewhere in the npm registry or github if a package has automatic updates, but I'd definitely take a look at their github repo and see if the package has recent updates and contributors.'' - R26}

\conclusion{Finding 3: Practitioners are likely to incorporate our findings into their dependency management practices, but not all attributes are readily accessible to downstream dependents.}

%
%
\section{Implications}
\label{sec:Implications}
We present actionable implications for practitioners and researchers.

\noindent\textbf{Implications for Practitioners \& Maintainers:}

Delay in the adoption of fixes is the main reason packages remain vulnerable (RQ1). Almost all vulnerabilities in our dataset are fixed within a day after public disclosure, but it takes an average of 6 months for the fix to be adopted by downstream dependents (RQ1). There are many tools available for notifying developers of vulnerabilities in direct dependencies. For example, Dependabot is an open source tool can identify vulnerable dependencies and raise a pull request to adopt the fix \cite{dependabot}. SonarQube \cite{sonarqube} is an enterprise-ready alternative that can identify vulnerable dependencies using its dependency check plugin \cite{sonarqube_plugin}. However, developers must also be aware of the \textit{transitive} influence of upstream packages on their project. Specifically, \textbf{developers should maintain an inventory of their entire dependency tree (direct and transitive) and periodically analyze their transitive dependencies for vulnerabilities}. As one of the respondents in our survey (RQ3) highlighted: ``Maintain some sort of SBOM [Software Bill of Materials] and monitor for transitive dependencies with known CVEs''. In addition, \textbf{developers can use our proposed set of attributes (RQ2) as dependency selection criteria, because the consequences of bad dependency management trickles downstream}.

Multiple respondents in our practitioner survey (RQ3) cited the lack of accessibility as a reason for not using some of the suggested attributes. The npm registry uses badges to display certain metadata about a package such as the number of downloads and test coverage. \textbf{Package maintainers should expand the use of badges to include information sought after by potential downstream dependents to aid their dependency selection process}. A good example would be to display information regarding dependency update strategies (RQ2) such as the number of pinned dependencies. This functionality is currently available with third-party tools such as Dependency Sniffer \cite{dependencysniffer}. Additionally, \textbf{Vulnerability databases should add information regarding vulnerable dependencies for a package to provide a more holistic depiction of risk for potential dependents}.


\noindent\textbf{Implications for Researchers:}

We use a set of package attributes to model how fast the package will respond to a vulnerability fix in their dependencies (RQ2). However, as highlighted by our survey respondents (RQ3), socio-technical factors play a significant role in the responsiveness of a package to vulnerable dependencies. When asked about identifying packages with a speedy response to vulnerability fixes, one respondent stated: ``For me, looking at who is behind the package also helps. Is it a college kid pushing his first npm package or is it a new project from Microsoft?''. Another respondent said: ``[...] packages owned by large corps (e.g. Microsoft, Meta...) will be faster in checking for these issues even before they have many users''. \textbf{Future research should investigate the organizational practices that influence the responsiveness to software vulnerabilities}, especially in the case of open source projects tied to large organizations.

We found that the dependency update strategy is one of the important indicators for predicting the response speed to vulnerability fixes (RQ2). An important reason why developers are hesitant to freely their dependencies is the fear of breaking changes. For example, one of the respondents in our survey (RQ3) said: ``[...] when packages update, and break code (which can happen even for patch releases, even though it shouldn't)[...]''. Another respondent stated: ``For major changes it is not technically possible to allow automatic updates since the major updates includes breaking changes for other dependencies''. Although Semantic Versioning was proposed to alleviate such issues, it is still not adopted by all packages in the npm ecosystem \cite{decan2019package}. In fact, it is not uncommon for developers to downgrade to a previous version, following a seemingly compatible update \cite{cogo2019empirical}. Different ecosystems favor different practices and policies. Different packaging ecosystems can also have a differing levels of Semantic Versioning adoption \cite{decan2019empirical, dietrich2019dependency, li2023large} and different cultural habits \cite{bogart2016break, values2020}. \textbf{Future research should study the ecosystem-specific attributes and policies that indicate responsive packages}, especially across packaging ecosystems with diverse policies.


%
\section{Related Work}
\label{sec:Related Work}
In addition to the studies cited throughout the paper, this section describes the key research works that study vulnerable dependencies in the npm ecosystem and approaches for selecting dependencies.

\noindent\textbf{Vulnerable dependencies:}

Decan et al. \cite{decan2018impact} explored the impact of security vulnerabilities in the npm ecosystem. They found that the number of vulnerabilities in the ecosystem are on the rise but most vulnerabilities are fixed before they are publicly disclosed. They observed that a large fraction of packages do not immediately adopt the fix released by the upstream package, leaving them vulnerable despite the availability of the fix. The authors also highlighted that from a package user's perspective, there is no difference between being directly exposed to a vulnerability or being exposed to a vulnerability through a dependency.

Chinthanet et al. \cite{chinthanet2021lags} studied the release and adoption of vulnerability fixes in the npm ecosystem. They found that vulnerability fixes are not always released as a patch, but often bundled into other release types. Additionally, the majority of the commits in the majority of the fixing releases are not related to the security vulnerability. The authors found that even when the fix is released as a patch, the direct dependent package often releases the fixed package as a minor or major version, meaning as we travel downstream in the dependency chain, relying on patch-only fixes is increasingly ineffective. They also observed that the type of the release and the severity of the vulnerability influence the propagation of the fix across the ecosystem.

Alfadel et al. \cite{alfadel2023discoverability} conducted an empirical study on Node.js applications to analyze the discoverability of npm vulnerabilities. They found that 67\% of applications have at least one vulnerable dependency. The main reason for the existence of publicly disclosed vulnerable dependencies in projects was the refusal to update the dependency to a newer version. In half of the applications studied by the authors, exposure to publicly disclosed vulnerabilities persists for more than 3 months. Additionally, the authors found that the majority of the projects (77\%) are infected by a small subset of 5 vulnerability types.

Zimmermann et al. \cite{zimmermann2019small} studied the potential of individual packages and package maintainers to threaten the security of the npm ecosystem. They found that installing an average npm package creates an implicit trust on 79 unique packages and 39 unique maintainers. Additionally, they found that the top 5 packages in npm are used by more than 100,000 downstream dependents, which makes such packages a primary target for attackers. The authors cited characteristics of the npm ecosystem such as heavy reuse, micropackages (small packages with few lines of code) and an open publishing model as potential security threats.

Zerouali et al. \cite{zerouali2022impact} empirically analyze the impact of vulnerabilities on transitive dependents in the npm and RubyGems ecosystems. They observe that it takes up to 7 years to disclose half of the lingering vulnerabilities in the npm ecosystem. They also found that more than 15\% of the latest dependent releases in npm are exposed to vulnerabilities from direct dependencies and 36.5\% of the latest releases are exposed to vulnerabilities from transitive dependencies. The authors found that the number vulnerabilities from transitive dependencies decrease as you go deeper along the dependency chain but vulnerabilities are still existent at the deepest levels. 

\noindent\textbf{Dependency selection:}

Suhaib et al. \cite{mujahid2023characteristics} examined the characteristics of highly selected packages in the npm ecosystem. Through their qualitative analysis, they found that developers gravitate towards popular packages that have adequate documentation and are generally free from vulnerabilities. Their quantitative analysis of more than 2,500 packages confirmed their observations. The authors highlighted that developers should carefully consider the attributes of a package before adding it as a dependency of their project. They further mentioned that package maintainers should strive to make such attributes more accessible to their downstream dependents. 

Vargas et al. \cite{larios2020selecting} studied the technical, human and economic factors considered by practitioners when selecting dependencies. The authors present the importance of release characteristics such as active maintenance and stability of packages in the dependency selection process, but underline the lack of a standard means of measuring such factors. The authors also observe disagreements between developers on the correct approach to selecting dependencies. They highlight the need to move away from ad-hoc decision-making towards a more systematic means of identifying suitable packages.

Pashchenko et al. \cite{pashchenko2020qualitative} conducted 25 interviews to understand how developers select packages. The found that developers consider the community support of a package as an important factor when selecting dependencies. They also observed that developers have different dependency management practices but generally regard vulnerabilities as an important factor in dependency management decisions. Developers expressed frustration with packages with packages with a high number of dependencies due to the lack of control over transitive dependencies.

\textit{Our study} builds on the previous works regarding vulnerabilities and dependency selection in the npm ecosystem by proposing empirically extracted dependency management practices that are associated with a faster response to vulnerability fixes. In addition to providing an in-depth analysis of how certain package attributes and behaviors influence the adoption of vulnerability fixes, our practitioner-aligned solution to selecting dependencies provides a means to mitigate the impact of vulnerabilities from transitive dependencies which are commonplace in the npm ecosystem.

\section{Threats to Validity}
\label{sec:Threats to Validity}
In this section, we discuss the threats to the validity of our study.\\
\textit{Threats to construct validity:} 
Threats to construct validity refer to the concern between the theory and the results of the study. 
In order to measure \textit{responsiveness}, we categorize the adoption delay into fast and slow classes based on a threshold of less than 2 days and more than 14 days. However, there is no consensus on what is defined as fast or slow.  We initially experimented with a multi-class model by distributing the delay into 4 classes. A response of 2 days or less was classified as fast; a response of more than 2 days but less than 2 weeks was classified as acceptable; a response between 2 weeks and 3 months was classified as mediocre and a response later than 3 months was classified as slow. However, the fast and slow classes combined made up over 96\% of our class distribution, indicating that we are in fact dealing with a binary classification problem. We then conducted a sensitivity analysis to ensure minor changes in our threshold does not translate into a considerable change in our class distribution.  Reducing the threshold from 48 hours to 0 hours decreases the distribution of the Fast class from 63.7\% to 63.6\% (less than 1\% change). Increasing the threshold from 48 hours to 30 days increases the distribution of the Fast class from 63.7\% to 64.2\% (less than 1\% change). When identifying \textit{exposure to vulnerabilities} from dependencies, we assume all dependencies in a package are fetched from the npm package manager. In reality, developers can install a dependency from any source (e.g. GitHub). The problem with considering sources outside of the official package registry (npm) is that there is no way to extract the list of dependents for an ad-hoc package hosted on the web. There is also no way to guarantee what is installed by the package manager as the contents of the hosted package can change at any time.

A vulnerability fix can be released as a major, minor or patch version (Section~\ref{sec:Background}). Since we have the information for package versions and vulnerability metadata, we can compare the version number of the fixing release (r) with the version number of the release right before the fix (r-1) to evaluate the type of fixing release. 

\textit{Threats to internal validity:} 
Threats to internal validity refer to the concerns that are internal to the study such as experimenter bias and errors. We use 9 features that we believe can serve as indicators of the response to vulnerabilities. There could be additional indicators that are not captured (or not feasible) using our feature set. One example is the experience of the development team for a package, which can influence how they respond to vulnerabilities regardless of the package attributes (as hinted in the responses of RQ3). We do not claim our collection of features to be an exhaustive list of all of the relevant characteristics and behaviors for predicting the adoption of vulnerability fixes. However, as can be seen in Section~\ref{sec:Results}, our model model has a high capability of predicting the response to vulnerabilities. In order to extract the release type of the vulnerability fix in RQ1, we compare the version of the fixing release against the version of the release right before the fix. However, some packages may have multiple simultaneous release streams which means the chronological order of release may not align with the numerical order. The libraries.io dataset used in this study dates to January 2020. Collecting the metadata for an entire ecosystem from scratch requires great effort but is also prone to errors. The libraries.io dataset has been used in multiple prior studies \cite{decan2021lost, decan2019package, zerouali2019diversity, zerouali2022impact} and its accuracy has been verified by other researchers \cite{decan2019empirical}. Additionally, we study vulnerabilities that have been discovered, disclosed, fixed and propagated across the ecosystem. We are more interested in the dynamics of dependency management practices for vulnerability mitigation, rather than the response to the latest vulnerabilities.

\textit{Threats to external validity:}
Threats to external validity concern the generalization of our findings. The methodology for identifying package attributes that indicate a fast response to vulnerabilities is applicable to other ecosystems. However, the scope of our findings is focused on the npm ecosystem. Therefore, our results may not be applicable to other software packaging ecosystems, especially if they follow dependency guidelines that are considerably different than npm. We conduct a survey to understand the developer's perception on our findings. While our response rate of 59\% is considerably higher than the the usual rate in software engineering surveys based on questionnaires \cite{singer2008software}, having more respondents may influence our understanding on practitioner perspectives.

\section{Conclusion}
\label{sec:Conclusion}
The objective of our study was to propose a means to mitigate vulnerabilities in transitive dependencies. We curated a dataset of 450 vulnerabilities and over 200,000 unique dependents that are exposed to these vulnerabilities through their dependencies. We use 9 features to train a model that predicts the adoption speed of vulnerability fixes. We found that packages that younger packages that release more often, favor non-restrictive update strategies and have less dependencies are faster in adopting vulnerability fixes. We also conducted a survey of 67 industry practitioners to obtain their perception on our findings. We found that the experience of practitioners generally align with our proposed set of features. However, many were not aware of the importance of downstream dependency decisions in mitigating vulnerabilities. Previous research has frequently suggested that practitioners need to be wary of the risk of vulnerabilities from transitive dependencies. Developers can use our findings to incorporate the mitigation of vulnerabilities from transitive dependencies in their dependency selection criteria.

\bibliographystyle{abbrv}
\balance
\bibliography{main-base}

\end{document}